\documentclass{ws-ijmpc}
\usepackage[super,compress]{cite}
\usepackage{xcolor}

\begin{document}

\markboth{L. N. Carenza {\it et al.}}
{Dynamically asymmetric and bicontinuous morphologies in 
active emulsions}

\catchline{}{}{}{}{}

\title{Dynamically asymmetric and bicontinuous morphologies in 
active emulsions}

\author{Livio Nicola Carenza
}

\address{Dipartimento  di  Fisica,  Universit\`a  degli  Studi  di  Bari  and  INFN,
Sezione  di  Bari,  via  Amendola  173,  70126 Bari,  Italy \\
livio.carenza@ba.infn.it}

\author{Giuseppe Gonnella}
\address{Dipartimento  di  Fisica,  Universit\`a  degli  Studi  di  Bari  and  INFN,
Sezione  di  Bari,  via  Amendola  173,  70126 Bari,  Italy \\
giuseppe.gonnella@ba.infn.it}

\author{Antonio Lamura}
\address{Istituto Applicazioni Calcolo, CNR, Via Amendola 122/D, 70126 Bari, Italy \\
antonio.lamura@cnr.it}

\author{Giuseppe Negro}
\address{Dipartimento  di  Fisica,  Universit\`a  degli  Studi  di  Bari  and  INFN,
Sezione  di  Bari,  via  Amendola  173,  70126 Bari,  Italy \\
giuseppe.negro@ba.infn.it}

\maketitle

\begin{history}
\received{Day Month Year}
\revised{Day Month Year}
\end{history}

\begin{abstract}
The morphology of a mixture made of a polar active gel immersed in an 
isotropic passive fluid is studied numerically. Lattice Boltzmann method is adopted 
to solve the Navier-Stokes equation and
coupled to a finite-difference scheme used to integrate the dynamic equations of the
concentration 
and of the polarization of the active component.
By varying the relative amounts of the mixture phases, different structures can be
observed. In the contractile case, at moderate values of activity, elongated structures are
formed when the active component is less abundant, while a dynamic emulsion of passive droplets in an 
active matrix is obtained for symmetric composition. When the active component is extensile, 
aster-like rotating droplets and a phase-separated pattern appear for asymmetric and symmetric
mixtures, respectively. The relevance of space dimensions in the overall morphology is shown 
by studying
the system in three dimensions in the case of extensile
asymmetric mixtures where interconnected tube-like structures span the whole system.

\end{abstract}

\ccode{PACS Nos.: xxx, xxx}

\section{Introduction}

The ability of some systems to perform autonomous motion when energy 
is provided either by 
external sources or by internal environment, is a fascinating and hot topic 
in recent research in physics \cite{rama2010}.  
The plethora of emerging phenomena from self-assembly to spontaneous flow, 
from bacterial turbulence to collective motion 
\cite{marc2013,elge2015,bech2016} opened the way to explore new phenomena
in the so-called active matter
and to set up materials with prescribed properties.

The study of single-component active systems is quite extended while
the case of mixtures with active and passive phases has been 
poorly investigated. Brownian simulations of active and passive particles
enlightened the role of activity in separating the two species 
\cite{mccandlish2012spontaneous,PhysRevLett.119.268002}. Binary fluids were
showed to have unstable active-passive interface due to the active 
component \cite{tjhu2012,blow2014biphasic}. The model for emulsification of
an active component in a passive fluid due to the presence
of surfactant, has been recently 
introduced and studied for different compositions \cite{bonelli2018,negro2018}. 

The aforementioned model generalizes the well established theory of
active gels 
\cite{joanny2009active,prost2015active} to describe a polar gel 
with extensile or contractile activity dispersed 
in an isotropic passive fluid. It was shown that intriguing patterns can be tuned 
by varying both the magnitude of the activity and the relative amounts of the two
components. In the present paper we illustrate the morphologies that can be 
obtained for moderate values of activity when the active phase is equal to, smaller or larger 
than $50\%$ in two-dimensional systems.
When the active motion is puller-like, deformed patterns are
observed when the active component is less abundant. A dynamic emulsion of passive droplets in an 
active structure can be triggered for symmetric composition. When the active component is pusher-like, 
droplets, resembling asters, and a phase-separated pattern appear for asymmetric and symmetric
mixtures, respectively. We also show for the first time how a system in three dimensions,
in the case of extensile asymmetric mixtures, allows to build 
interconnected tube-like structures spanning the whole system. 

The equilibrium properties of the model are derived from a proper free-energy functional. The
phenomenological dynamic equations for the concentration and for the local orientation of the active
phase are derived from the free energy and coupled to the Navier-Stokes equation for the overall fluid.
The equations are numerically solved by using a hybrid lattice Boltzmann method 
\cite{tiri2009,cate2009,tjhu2012}. 

\section{Model and Methods}

We outline here the hydrodynamic model and the method used to perform this numerical study.
We consider a fluid mixture made of an active polar phase and of a solvent, with overall mass 
density $\rho(\bm{r},t)$ 
and velocity $\bm{v}(\bm{r},t)$. We describe the relative concentration of the two phases with 
a scalar field $\phi(\bm{r},t)$ and introduce the vector field $\bm{P}(\bm{r},t)$ defining the local 
average orientation of the active material.
The evolution of the system is governed by the following set of equations in the 
limit of incompressible flow:
\begin{align}
\rho \left(\partial_t + \bm{v}\cdot \nabla\right) \bm{v} = - \nabla P 
+ \nabla \cdot \left( \underline{\underline{\sigma}}^{passive} 
+\underline{\underline{\sigma}}^{active} \right), \label{eqn:NS}\\
\frac{\partial \phi}{\partial t}+\nabla\cdot\left(\phi\mathbf{v}\right)
=M \nabla^2 \mu, \label{eqn:convection_diffusion}\\
\frac{\partial\mathbf{P}}{\partial t}+\left(\mathbf{v}\cdot\nabla\right)\mathbf{P}
=-\underline{\underline{\Omega}}\cdot\mathbf{P}
+\xi\underline{\underline{D}}\cdot\mathbf{P}-\frac{\bm{h}}{\Gamma} \label{eqn:beris_edwards}.
\end{align}

Here the first equation is the incompressible Navier-Stokes equation, where $P$ is the ideal 
gas pressure, while the stress tensor has been divided in a passive contribution 
$\underline{\underline{\sigma}}^{passive}$, given by the conserved momentum current, and a 
phenomenological active contribution $\underline{\underline{\sigma}}^{active}$ \cite{beris1994}.
The second equation rules the evolution of the concentration field. Being globally conserved, 
this can be described by a convection-diffusion equation where the chemical potential 
$\mu=\frac{\delta \mathcal{F}}{\delta \phi}$ is derived from an appropriate free energy 
$\mathcal{F}\left[\phi, \bm{P}\right]$ (to be defined in the following), and $M$ is the mobility parameter.
The evolution of the polarization field is governed by an Ericksen-Leslie equation, 
adapted for the treatment of a vector order parameter. Here 
$\bm{h}=-\frac{\delta \mathcal{F}}{\delta \bm{P}}$ is the molecular field, $\Gamma$ is 
the rotational viscosity, $\underline{\underline{D}}=\frac{1}{2}(\nabla \bm{v} + \nabla \bm{v}^T)$ is the 
symmetric deformation rate tensor, $\underline{\underline{\Omega}}=\frac{1}{2}(\nabla \bm{v} - \nabla \bm{v}^T)$
is the vorticity 
tensor; and $\xi$ 
is a parameter that controls the aspect ratio of active particles (positive for rod-like particles 
and negative for disk-like ones).
To define the equilibrium properties of our system we made use of the Landau-Brazovskii theory 
\cite{braz}
 coupled to the distortion free-energy of a polar liquid crystal in the single-constant limit:
\begin{equation}
\label{eqn:freeE}
\begin{split}
\mathcal{F} \left[ \phi, \bm{P} \right] = \int d \bm{r} \left[ \dfrac{a}{4 \phi_{cr}^2} \phi^2 (\phi - \phi_0)^2 
+ \dfrac{k_\phi}{2} (\nabla \phi)^2 + \dfrac{c}{4} (\nabla^2 \phi)^2 \right. \\ \left. 
-\dfrac{\alpha}{2} \dfrac{(\phi-\phi_{cr})}{\phi_{cr}} \bm{P}^2 + \dfrac{\alpha}{4} \bm{P}^4 
+ \dfrac{k_P}{2} (\nabla \bm{P})^2 + \beta \bm{P} \cdot \nabla \phi \right] .
\end{split}
\end{equation}
By choosing $a>0$, the separation of the two phases is favored with equilibrium values $0$ and
$\phi_0$, while the gradient terms 
rule the intensity of the interfacial tension. Notice that by choosing $k_\phi<0$, interfaces 
are likely to form, so that the Brazovskii constant $c$ must be positive to guarantee 
thermodynamic stability. We will name as \emph{active} those regions where 
$\phi>\phi_{cr}\equiv \phi_0/2$.
The terms in $\bm{P}$ in Eq.~\ref{eqn:freeE} define the equilibrium properties for the 
polarization field, that is confined in those regions where $\phi>\phi_{cr}$, for positive 
values of $\alpha$. The energy cost for the elastic deformations is paid by the gradient 
term, while the coupling between the polarization and the concentration field defines the 
properties of anchoring of the vector field at interfaces. In particular if $\beta>0$ then the 
polarization at interfaces will point towards passive regions of the mixture.
This free energy has a transition line from the ferromagnetic phase to the lamellar phase 
at $a_{cr}=\frac{k_{\phi}^2}{4c} + \frac{\beta^2}{k_P}$ when the two components
are symmetric. In the case of asymmetric
composition stable droplets of the minority phase are observed 
\cite{kk,Gompper_critic,gonnella1998}. This is equivalent to disperse a suitable amount of surfactant in the mixture favoring emulsification of the two phases.

In order to make the description of the model complete, we explicitly write the various 
contributions to stress tensor, appearing in Eq.~\ref{eqn:NS}. For what concerns the 
passive part of the stress tensor, this includes the viscous contributions 
$\underline{\underline{\sigma}}^{viscous}$ and the conserved momentum current stemming from 
the free energy, including binary mixture and polarization contributions. Thus we write:
\begin{equation}
\underline{\underline{\sigma}}^{passive}=\underline{\underline{\sigma}}^{viscous}
+\underline{\underline{\sigma}}^{polar} 
+ \underline{\underline{\sigma}}^{binary}
\end{equation}
where $\sigma^{vis}_{\alpha \beta} = \eta (\partial_\alpha v_\beta + \partial_\beta v_\alpha )$ is the usual viscous tensor stress, $\eta$ being the shear viscosity;  $\sigma^{pol}_{\alpha \beta} = \dfrac{1}{2} (P_\alpha h_\beta - P_\beta h_\alpha) 
- \dfrac{\xi}{2}(P_\alpha h_\beta + P_\beta h_\alpha) - k_P \partial_\alpha P_\gamma \partial_{\beta} P_\gamma$ 
comes from liquid crystal theory \cite{beris1994}, 
$\xi$ determining whether the fluid is flow tumbling ($|\xi|<1$)
or flow aligning ($|\xi|>1$); $\sigma^{bin}_{\alpha \beta} = \left( f - \mu \phi \right) \delta_{\alpha \beta} 
- k_{\phi} \partial_\alpha \phi \partial_{\beta} \phi +c \left[ \partial_\alpha \phi \partial_\beta(\nabla^2 \phi ) 
+ \partial_\beta \phi \partial_\alpha (\nabla^2 \phi) \right] - \beta P_{\beta} \partial_{\alpha} \phi$ includes the interfacial stress, $f$ being the free energy density appearing in Eq.~\ref{eqn:freeE}.
Finally, $\sigma_{\alpha \beta}^{active} 
= -\zeta \phi \left( P_\alpha P_\beta - \frac{\bm{P}^2}{d} \delta_{\alpha \beta} \right)$ 
is a phenomenological term stemming from a coarse-graining 
procedure \cite{Simha,elsen}, 
assuming that polar objects are capable to do some work in their neighborings. 
Here $d$ denotes the space dimensions and $\zeta$ is the \emph{activity} which
allows to tune the energy supply due to the active component in the mixture. Its values are 
negative for contractile systems and positive for
extensile ones.

The phenomenological equations are solved numerically by using a hybrid lattice
Boltzmann (LB) method \cite{tiri2009,cate2009,gonnella2010,glt,Bonelli} 
where
a LB approach for the Navier-Stokes equation 
(see Ref.~\refcite{negro2018} for details) is coupled
with a finite-difference predictor-corrector method to integrate the equations
for the fields $\phi$ and $\bm{P}$. 
In $2d$, square lattices of size $L=256$ were adopted 
with periodic boundary conditions.

We will show here, for the first time, results from simulation in $3d$ space. 
Due to high-demanding computational costs in term of memory allocation and simulation times, we adopted a parallel approach by means of an MPI protocol, deviding the computational grid into slices and associating each slice to a single computational unit. Lattice Boltzmann solvers perform only local operation on data, and it is therefore optimal for parallelization. Computation of derivatives on the edge of subdomains has been solved by means of the \emph{ghost cells method} \cite{MPI}.

The range of values of $\phi$ goes from $\phi \simeq 0$ (passive phase)
to $\phi \simeq 2$ (active phase). 
The parameters are set to $a= 4 \times 10^{-3}$, $k_{\phi}= -6 \times 10^{-3}$,
$c= 10^{-2}$, $\alpha= 10^{-3}$, $k_P= 10^{-2}$, $\beta= 10^{-2}$,
$\Gamma= 1$, $\xi= 1.1$, $\phi_0= 2$, and $\eta=1.67$.
In the following all the quantities are measured in lattice units.

Mixtures with different amounts of active and passive components were
considered denoting the compositions with the notation $x:(100-x)$ where
$x$ refers to the active part. Systems were initialized with random 
configurations of polarization and concentration and let evolve in the
presence of activity.

\section{Results}
In the following we will give a characterization of the system in the case
of contractile ($\zeta<0$) and extensile ($\zeta>0$) activity for different
composition of the mixture.
In order to better appreciate the role played by activity we start by
discussing the equilibrium configuration in the passive limit ($\zeta=0$).
For symmetric preparation of the mixture the system described by the
free-energy functional of Eq. \eqref{eqn:freeE} sets into a lamellar phase with some dislocations in the pattern.
For sufficiently asymmetric preparations of the mixture (the critical concentration is around $35:65$ for the choice of parameters used in the present study but it generally depends on the surface tension at interface) the system sets into a stable hexatic array of droplets of the minority phase in a majority background matrix. Domains are thus prevented from merging due to the presence of surfactant in the mixture as discussed in the previous Section.
\begin{figure}
\centering
\includegraphics[width=0.45\textwidth]{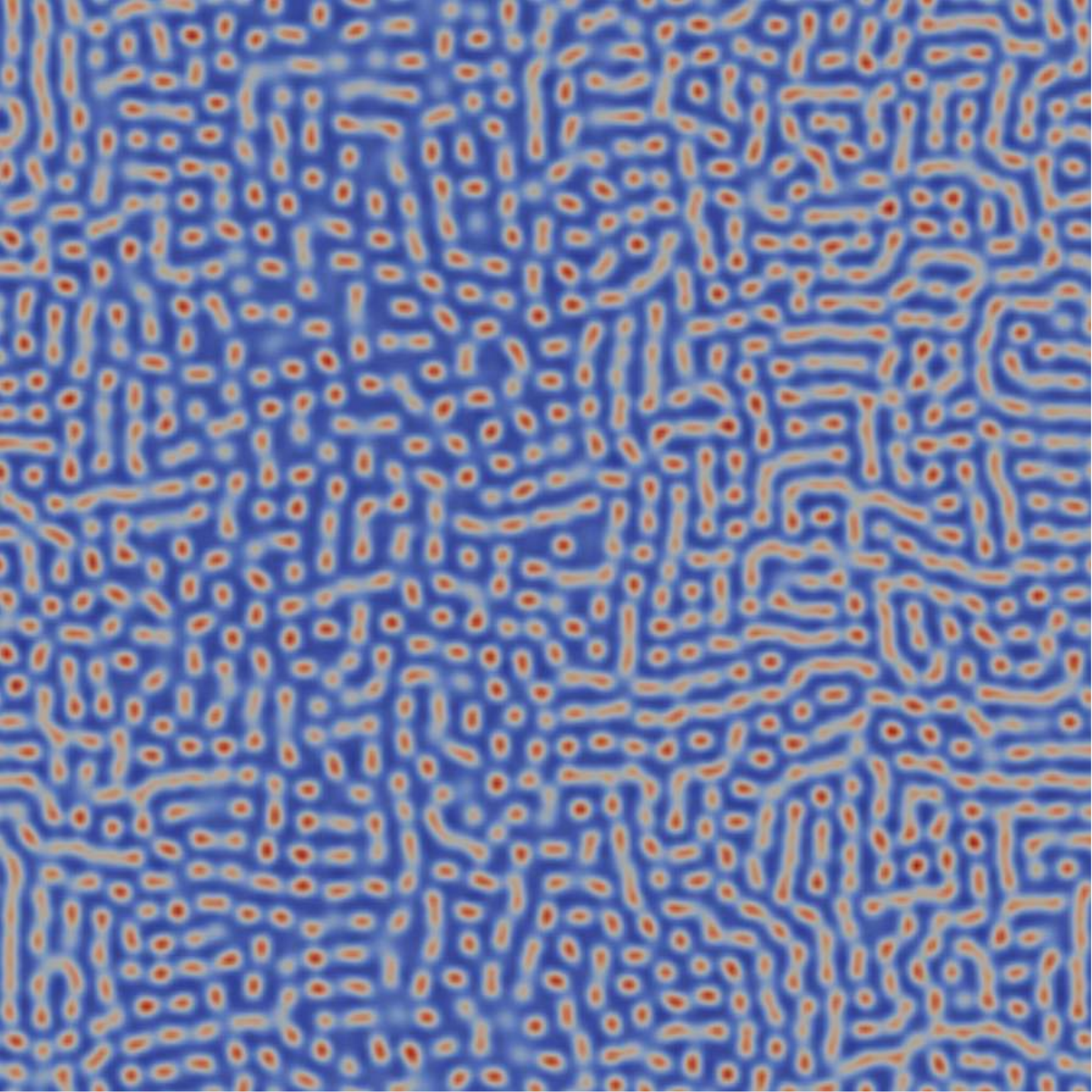} \  
\includegraphics[width=0.45\textwidth]{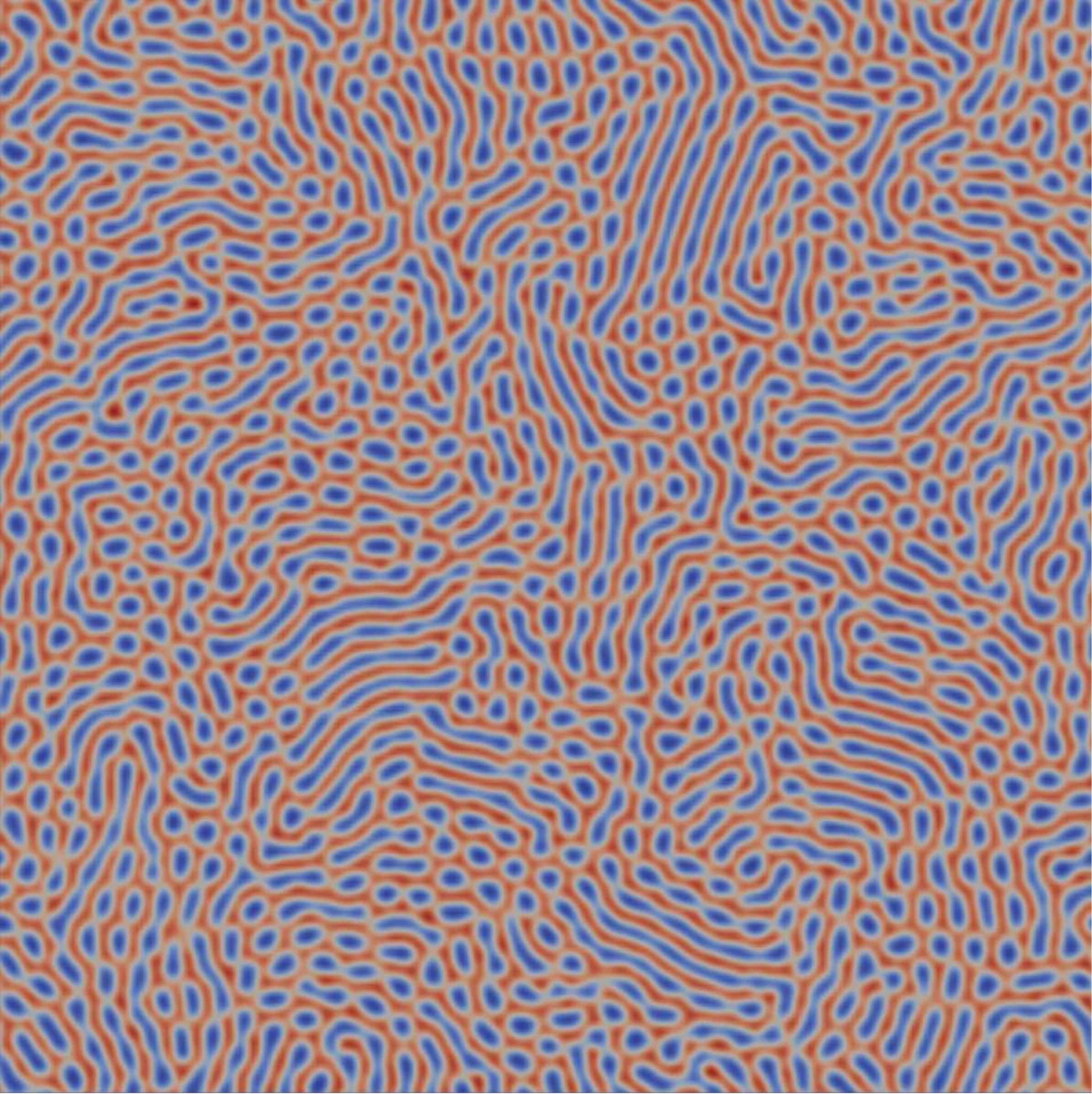}
\caption{Contour plots of $\phi$ at non-equilibrium steady states 
of contractile 10:90 (left) and 50:50 (right) mixtures rispectively at activity $\zeta=-0.02$ and $\zeta=-0.01$.}
\end{figure}

The effect of the activity on the morphology of the system is of crucial relevance in both contractile and extensile case.
We start our discussion from the contractile case. When small values of $|\zeta|$ are used, patterns do not change with respect to the equilibrium configurations previously discussed. Nevertheless, if contractile activity is increased, self-generated flows greatly affect the overall morphology. In particular contractile systems are unstable under splay deformation of the polarization pattern, thus droplets in asymmetric mixtures tend to deform in the direction of the main flow, due to strong anchoring of polarization to interfaces, leading to reduction of effective surface tension.
If contractile activity is sufficiently strong, this is enough to push the system towards formation of bicontinuous structures of the two phases even in the  asymmetric preparations (see left panel in Fig. 1).
As asymmetry between active and passive phase is reduced, the mechanism leading to surface tension reduction remains unaltered; nevertheless the effects on morphology are yet different: polarization tends to reduce elastic splay stresses by formation of point defects. This, together with surface anchoring, leads the the formation of an emulsion of droplets of passive materials in an active background matrix. These amorphous configurations never set into a stable pattern due to continuous energy injection at small scales, so that passive droplets are dragged and deformed by self-generated flows.

\begin{figure}
\centering
\includegraphics[width=0.32\textwidth]{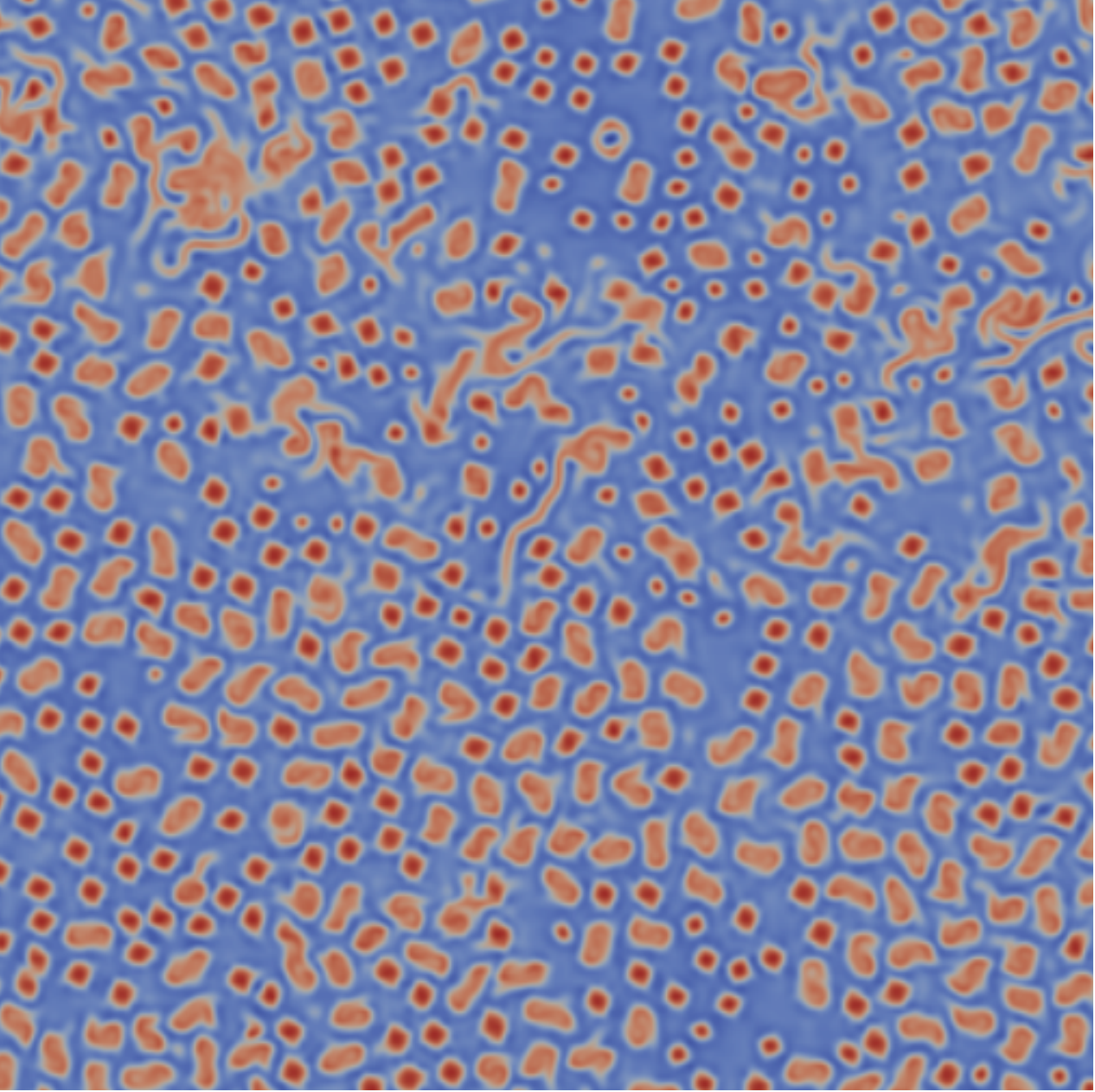} \  
\includegraphics[width=0.32\textwidth]{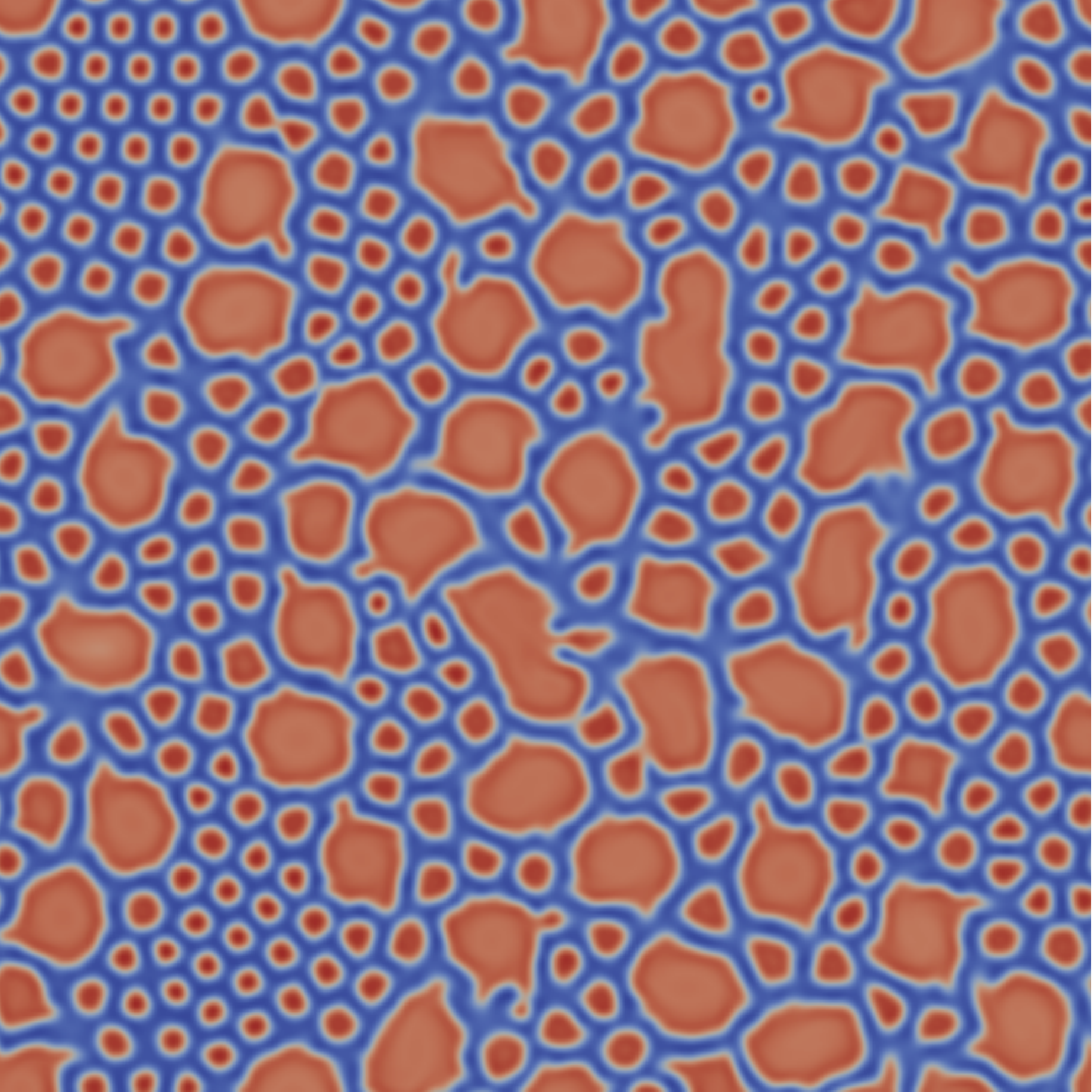} \ 
\includegraphics[width=0.32\textwidth]{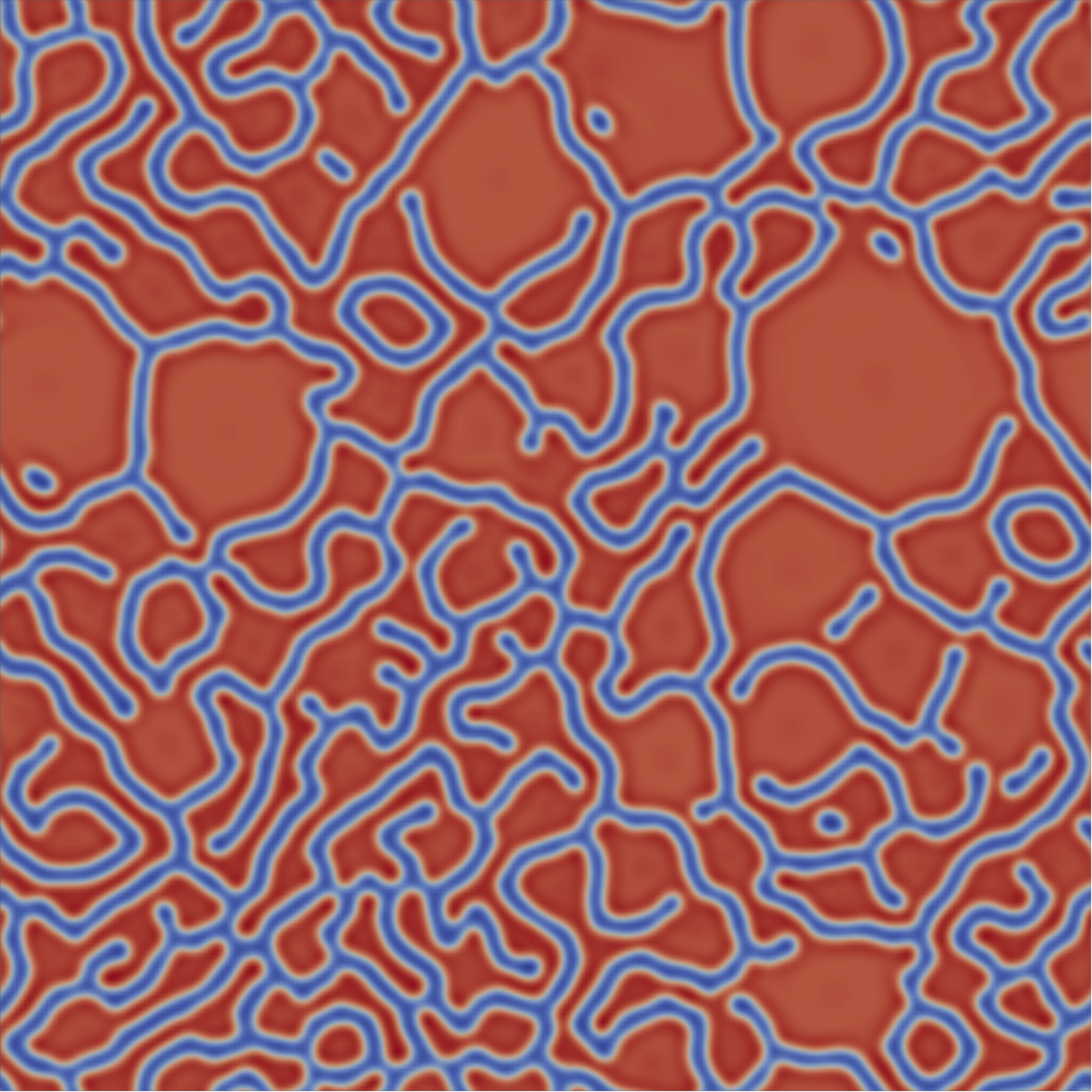}
\caption{Contour plots of $\phi$ at non-equilibrium steady states 
of extensile 10:90 (left) and 50:50 (center) and 75:25 (right) mixtures rispectively at activity $\zeta=0.008$, $\zeta=0.004$ and $\zeta=0.001$ .}
\end{figure}

If extensile activity is consdired we observe a net change in the overall behavior of the system. At low concentration of active material, the hexatic array of active droplets remains stable until a certain critical value is reached, depending on the total amount of the active component. Beyond this threshold droplets start to merge giving rise to big rotating droplets, driven by bending instability of polarization under extensile flows. 
As the fraction of active material is increased, keeping fixed the intensity of active doping $\zeta$, the system eventually undergoes demixing of the two phases giving rise to turbulent-like flows in the velocity pattern. 
Remarkably, in highly asymmetric mixtures, where the majority is active, we found a background passive matrix, despite it is the minority component of the mixture.

Even in $3d$ the effect of activity on the morphology is highly surprising. We consider here a highly asymmetric preparation ($10:90$) of the mixture. In the passive limit (see central panel of Fig. 3), analogously to what happen in the bidimensional case, an emulsion of droplets of the minority phase arrange in a cubic lattice with some dislocations in the arrangement. Droplets do not merge due to the presence of a suitable amount of surfactant, but if the minority phase is made active this ordered lattice structure is easily lost, whoever activity is contractile or extensile. In the former case, droplets for small values of $| \zeta |$ are first stretched by the flow, then as active doping is increased, droplets start to merge giving rise to tubular structures that span the system, creating a bicontinuous phase. The underlying mechanism is the same as descibed in the bidimensional case.
In the extensile case the morphological behavior is relatively different from its bidimensional counterpart. If in $2d$, bending instability in the polarization pattern acted as a source of vorticity, making droplets rotate, in $3d$ this is no more the case, since polarization is no more confined in a bidimensional plane and is free to rotate in space. Big amorphous ferromagnetic structures form in the system due to the increased effective surface tension. The thickening of active materials in such domains leads to formation of lighter structures in the rest of the system (see right panel in Fig. 3). 
The richness in morphology found in our preliminary $3d$ simulations deserves a further study, to methodically characterize the behavior of the system, while varying the intensity of active doping.

\begin{figure}[t]
\centering
\includegraphics[width=0.32\textwidth]{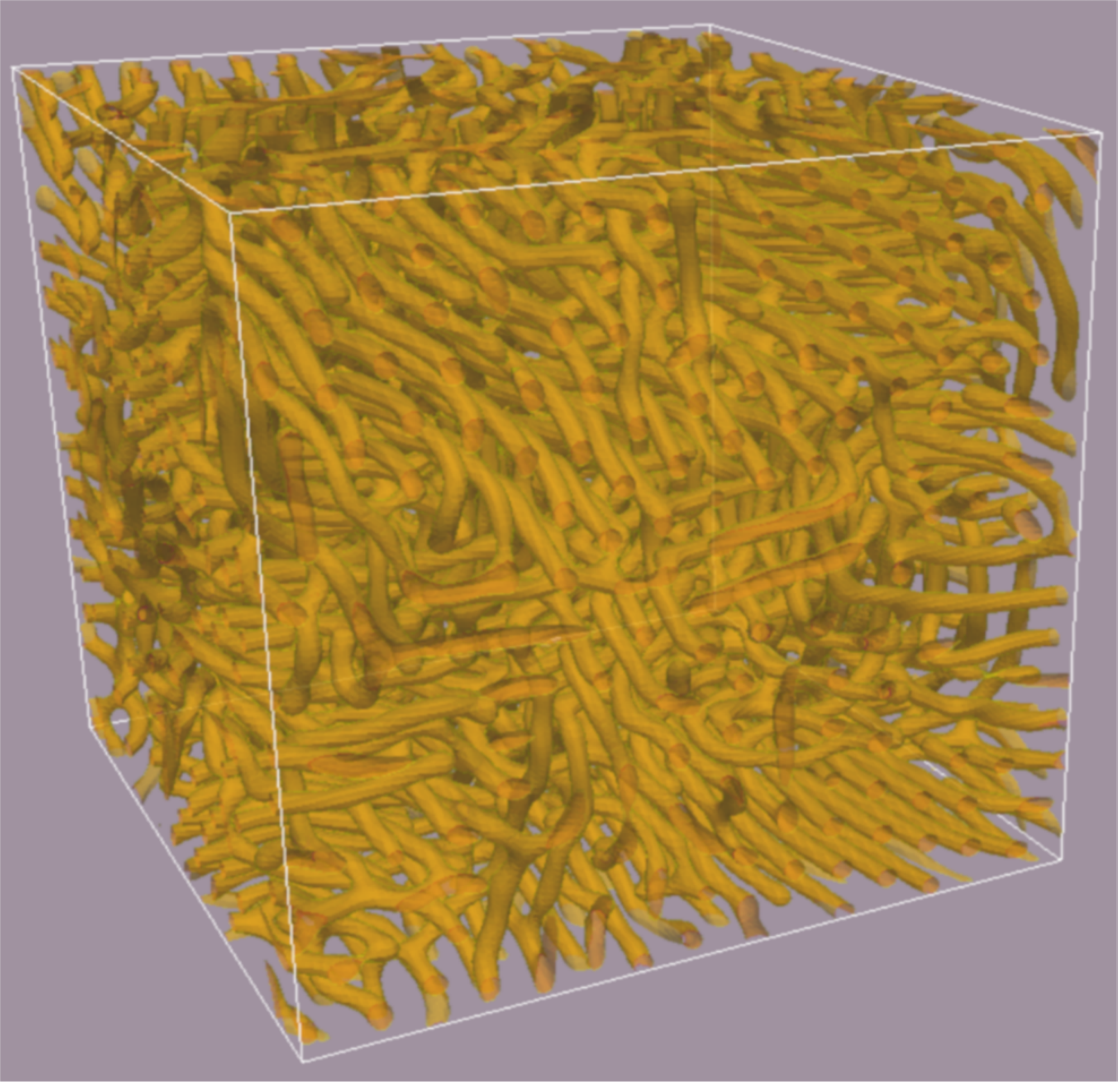} \  
\includegraphics[width=0.32\textwidth]{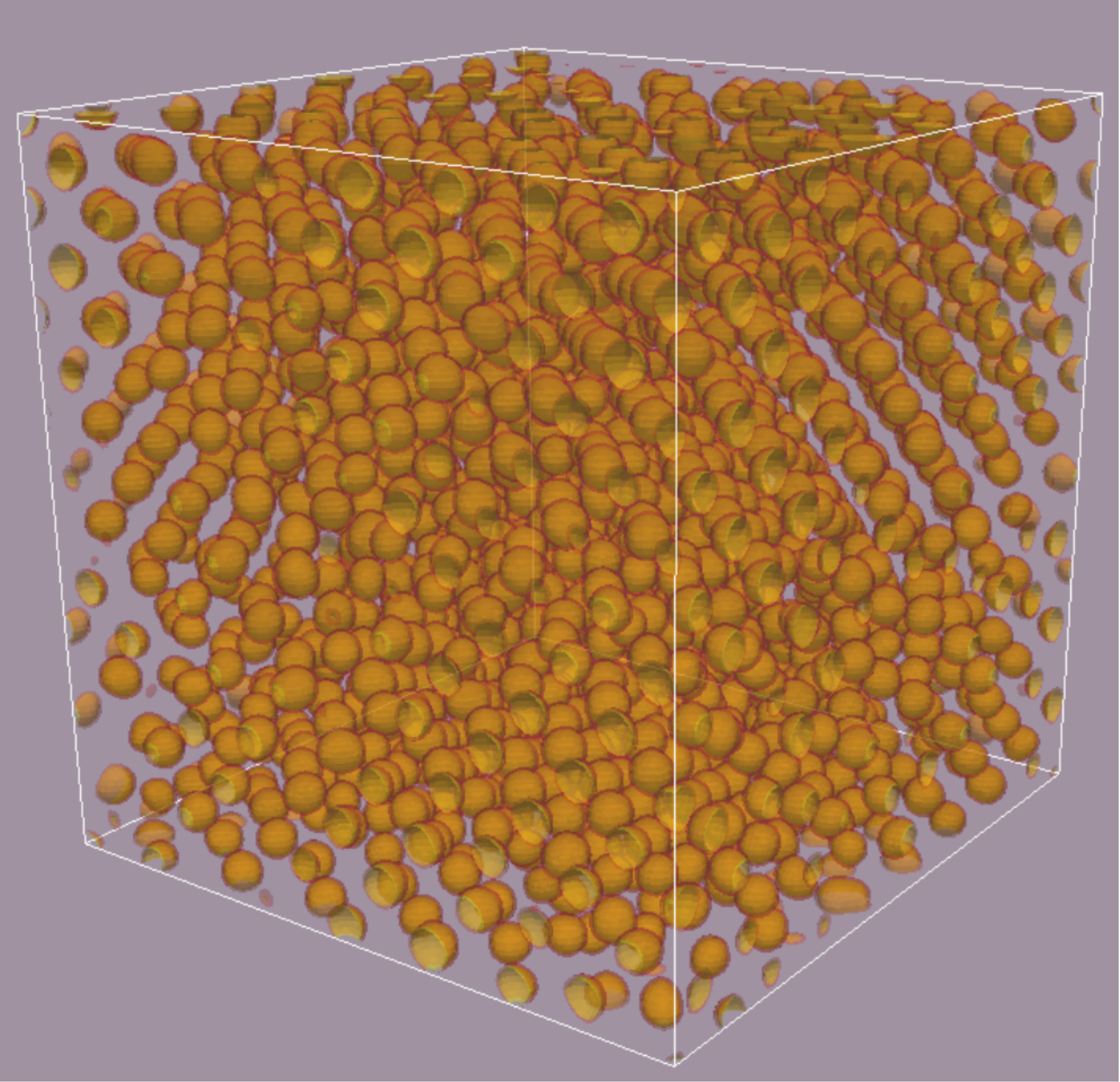} \
\includegraphics[width=0.32\textwidth]{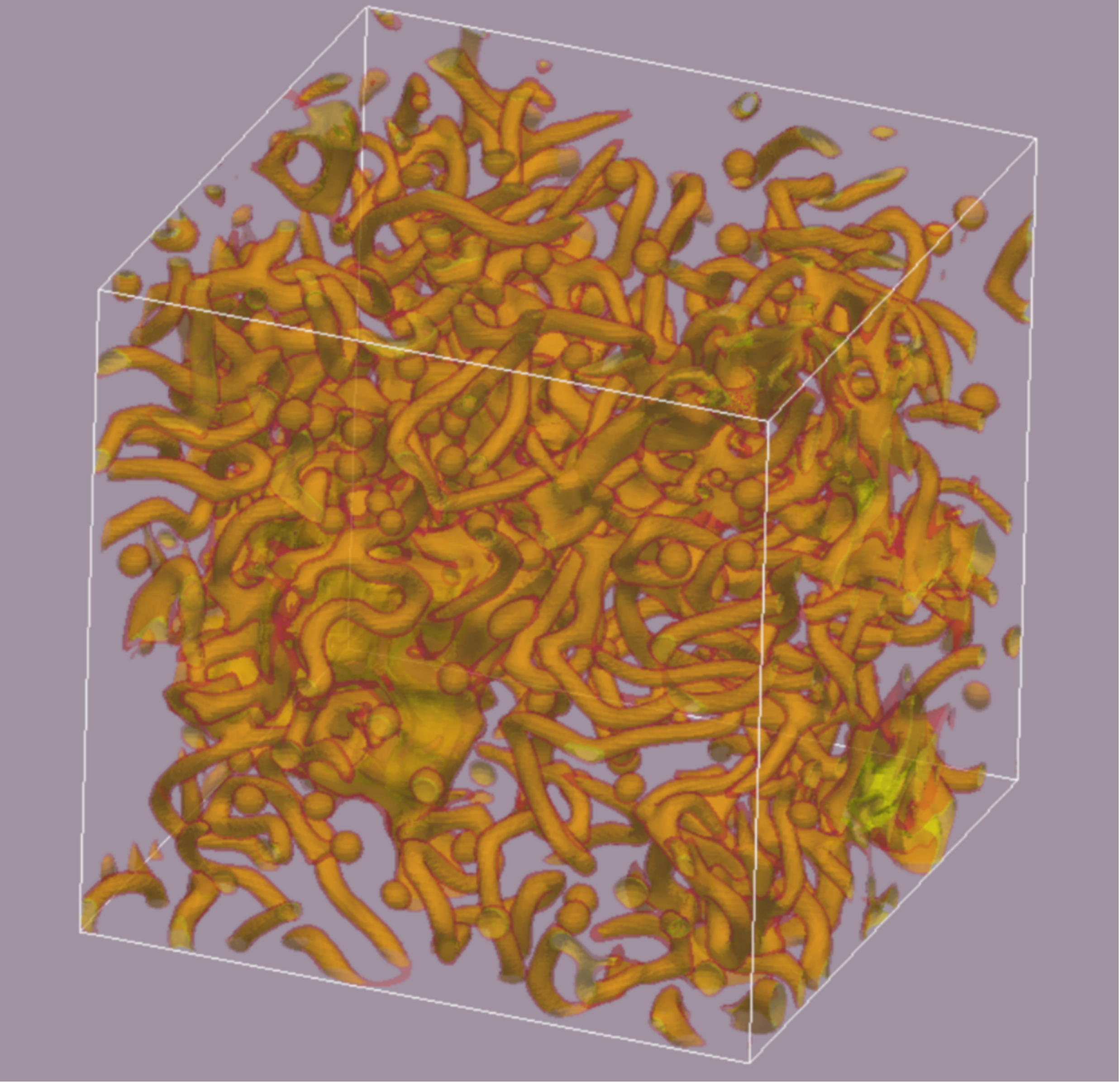} \
\caption{Isosurfaces at $\phi=1.5$ of concentration field $\phi$ in $3d$ systems. The central panel shows the relaxed configuration consisting in a cubic array of droplets. If contractile activity is switched on ($\zeta=-0.005$) a tubular network of active domains form a bicontinous phase that spans the whole system. In extensile systems ($\zeta=0.008$) ordered structures are lost in favor of amorphous and tubular shapes. }
\end{figure}

\section{Conclusions}

We have shown that an active polar gel embedded in an isotropic passive fluid can manifest
a wealth of structures by varying the relative amounts of the phases as well as the activity.
Our attention was focused on intermediate values of this latter parameter. 
Contractile emulsions were shown
to be characterized by the presence of elongated patterns that eventually form bicontinuous structures, even if the active phase is minority.
The same ratio of the two phases allows us to obtain emulsions of passive droplets in an active pattern.
For extensile emulsions flow patterns are responsible to produce much different morphological configurations, ranging from 
aster-like rotating droplets to phase-separated patterns, when increasing the amount of the active
component. Some preliminary results of three-dimensional systems have also been shown. These show that rich structures appear in the system while changing the kind and intensity of active doping,
suggesting that these systems as promising tools to fabricate smart materials able to self-assembly
with variable morphology.

\section*{Acknowledgments}

Simulations have been performed at Bari ReCaS e-Infrastructure funded by MIUR, Italy through the program PON Research
and Competitiveness 2007–2013 Call 254 Action I. We thank D. Marenduzzo, E. Orlandini and A. Tiribocchi for the very useful
discussions.

\bibliographystyle{ws-ijmpc}
\bibliography{refs2}

\end{document}